\documentstyle[amssymb,12pt,psfig]{elsart}
\begin{document}

%
%
%
%
\renewcommand {\AA}       {\mbox{A-A}}
\newcommand {\pp}       {\mbox{p-p}}
\newcommand {\pn}       {\mbox{p-n}}
\newcommand {\pA}       {\mbox{p-A}}
\newcommand {\pBe}      {\mbox{p-Be}}
\newcommand {\pCu}      {\mbox{p-Cu}}
\newcommand {\pAu}      {\mbox{p-Au}}
\newcommand {\pPb}      {\mbox{p-Pb}}
\newcommand {\SiAu}     {\mbox{Si-Au}}
\newcommand {\SiAl}     {\mbox{Si-Al}}
\newcommand {\SiA}      {\mbox{Si-A}}
\newcommand {\AuAu}     {\mbox{Au-Au}}
\newcommand {\PbPb}     {\mbox{Pb-Pb}}
\newcommand {\SuSu}     {\mbox{S-S}}
\newcommand {\SA}       {\mbox{S-A}}
\newcommand {\SAg}      {\mbox{S-Ag}}
\newcommand {\SPb}      {\mbox{S-Pb}}
\newcommand{\avg}[1]{\mbox{$\langle$ #1 $\rangle$}}
\newcommand {\AGeV}     {\mbox{A$ \cdot$GeV/c}}
\newcommand {\avgdeltay} {\mbox{$\langle \Delta y \rangle$}}
\newcommand {\AvgnuA}   {\mbox{$\langle \nu \rangle_A$}}
\newcommand {\Avgnungrey}   {\mbox{$\bar{\nu} (N_{\rm grey})$}}
\newcommand {\BBbar}    {\mbox{$\rm B\bar{B}$}}
\newcommand {\Cher}     {Cherenkov}
\newcommand {\dedx}     {\mbox{$dE/dx$}}
\newcommand {\avgdedx}  {\mbox{$\langle dE/dx \rangle$}}
\newcommand {\sumdedx}  {\mbox{$\sum \langle dE/dx \rangle$}}
\newcommand {\deltay}   {\mbox{$\Delta y$}}
\newcommand {\gammaconv}{\mbox{$\gamma \rightarrow e^{+}e^{-}$}}
\newcommand {\epem}     {\mbox{$e^+e^-$}}
\newcommand {\Et}       {\mbox{$E_t$}}
\newcommand {\fourpi}   {\mbox{$4\pi$}}
\newcommand {\Kp}       {\mbox{K$^+$}}
\newcommand {\Ktopi}    {\mbox{K$/\pi$}}
\newcommand {\Km}       {\mbox{K$^-$}}
\newcommand {\Ks}       {\mbox{K$_S$}}
\newcommand {\Kz}       {\mbox{K$^0$}}
\newcommand {\Kbar}     {\mbox{$\rm \bar{K}$}}
\newcommand {\Kzbar}    {\mbox{$\rm \bar{K^0}$}}
\newcommand {\KKbar}    {\mbox{${\rm K \bar{K}}$}}
\newcommand {\KpKbar}    {\mbox{${\rm K^+\bar{K}}$}}
\newcommand {\Ksdecay}  {\mbox{K$^0_S \rightarrow \pi^+\pi^-$}}
\newcommand {\Lam}      {\mbox{${\rm \Lambda}$}}
\newcommand {\LamminusLambar} {\mbox{${\Lambda - \bar{\Lambda}}$}}
\newcommand {\Lambar}   {\mbox{$\bar{\rm \Lambda}$}}
\newcommand {\LLbar}   {\mbox{$\Lambda \bar{\Lambda}$}}
\newcommand {\Lamdecay} {\mbox{${\rm \Lambda} \rightarrow p\pi^{-}$}}
\newcommand {\Minv}    {\mbox{$M_{\rm inv}$}}
\newcommand {\mperp}    {\mbox{$m_{\perp}$}}
\newcommand {\Ngrey}    {\mbox{$N_{\rm grey}$}}
\newcommand {\NN}       {\mbox{N-N}}
\newcommand {\NAB}      {\mbox{$\rm N^{AB}$}}
\newcommand {\Npart}    {\mbox{$N_{\rm part}$}}
\newcommand {\Npproj}    {\mbox{$N_{\rm pproj}$}}
\newcommand {\nubar}    {\mbox{$\bar{\nu}$}}
\newcommand {\nungrey}  {\mbox{$\nu(N_{\rm grey})$}}
\newcommand {\Nnu}      {\mbox{$\nu$}}
\newcommand {\Nnn}      {\mbox{$\rm N^{NN}$}}
\newcommand {\Npim}     {\mbox{$N_{\pi^-}$}}
\newcommand {\NkpPP}    {\mbox{$N_{\rm K^+}^{\rm pp}$}}
\newcommand {\NkpNN}    {\mbox{$N_{\rm K^+}^{\rm NN}$}}
\newcommand {\Om}       {\mbox{$\Omega^-$}}
\newcommand {\pbar}     {\mbox{$\bar{p}$}}
\newcommand {\pim}      {\mbox{$\pi^-$}}
\newcommand {\pion}     {\mbox{$\pi$}}
\newcommand {\pip}      {\mbox{$\pi^+$}}
\newcommand {\pippim}   {\mbox{$\pi^+\pi^-$}}
\newcommand {\pperp}    {\mbox{$p_\perp$}}
\newcommand {\ppim}     {\mbox{$p\pi^-$}}
\newcommand {\pz}    	{\mbox{$p_z$}}
\newcommand {\Rpart}    {\mbox{$\rm R_{part}$}}
\newcommand {\sig}      {\mbox{$\sigma$}}
\newcommand {\Sigz}      {\mbox{$\Sigma^0$}}
\newcommand {\sumpt}    {\mbox{$\sum p_\perp$}}
\newcommand {\sumpz}    {\mbox{$\sum p_z$}}
\newcommand {\rap}      {\mbox{$y$}}
\newcommand {\rhoN}     {\mbox{$\rho_{\rm N}$}}
\newcommand {\sigNN}    {\mbox{$\sigma_{\rm NN}$}}
\newcommand {\sqrts}    {\mbox{$\sqrt{s}$}}
\newcommand {\usec}     {\mbox{$\mu s$}}
\newcommand {\ylead}    {\mbox{$y_{\rm lead}$}}
\newcommand {\ynn}   	{\mbox{$y_{\rm NN}$}}
\newcommand {\dndy}     {\mbox{$dn/dy$}}
\newcommand {\mt}       {\mbox{$m_{\perp}$}}
\newcommand {\mtmo}     {\mbox{$m_{\perp}-m_0$}}
\newcommand {\VO}       {\mbox{$V_0$}}
\newcommand {\chisq}    {\mbox{$\chi^{2}$}}
\newcommand {\normchisq}{\mbox{$\chi^{2}_{V_0}/NDF$}}
\newcommand {\zerod}    {\mbox{$0^\circ$}}
\newcommand {\Xm}       {\mbox{$\Xi^-$}}

\newcommand{\vba}{\mbox{$\vec{b_{\rm A}}$}}
\newcommand{\vbb}{\mbox{$\vec{b_{\rm B}}$}}
\newcommand{\vb}{\mbox{$\vec{b}$}}
\newcommand{\Npnu}{\mbox{$N_{\rm p}(\nu)$}}
\newcommand{\AB}{\mbox{A-B}}
\newcommand{\bAB}{\mbox{$b_\AB$}}
\newcommand{\NLamPP}{\mbox{$N_\Lambda^{\rm pp}$}}
\newcommand{\NLam}{\mbox{$N_\Lambda$}}
\newcommand{\NLamNpart}{\mbox{$N_\Lambda(N_{\rm part})$}}
\newcommand{\Pnu}{\mbox{$P_{\rm p/t}(\nu)$}}
\newcommand{\PptNu}{\mbox{$P_{\rm p/t}(\nu)$}}
\newcommand{\bvec}{\mbox{$\vec{b}$}}
\newcommand{\bpv} {\mbox{$\vec{b_{\rm p}}$}}
\newcommand{\btv} {\mbox{$\vec{b_{\rm t}}$}}
\newcommand{\bt} {\mbox{$b_{\rm t}$}}
\newcommand{\bp} {\mbox{$b_{\rm p}$}}
\newcommand{\Rb}{\mbox{$R(b)$}}
\newcommand{\Npartb}{$\Npart(b)$}
\newcommand{\NKpPP}{\mbox{$N_{\rm K^+}^{\rm pp}$}}
\newcommand{\RNpart}{\mbox{$R(N_{\rm part})$}}
\newcommand{\nupart}{\mbox{$\nu_{\rm part}$}}

%
%
\begin{frontmatter}
\title{
Strangeness Enhancement in p-A Collisions: Consequences for the Interpretation of Strangeness Production in A-A Collisions
}
%
%
%
%
%
\author{B.A.~Cole$^{2}$}
\author{X.H.~Yang$^{2}$}
\author{V.~Cianciolo$^{6}$}
\author{A.~Frawley$^{3}$}
\author{E.P.~Hartouni$^{5}$}
\author{H.~Hiejima$^{2,\#}$}
\author{A.~Lebedev$^{4}$}
\author{N.~Maeda$^{3}$}
\author{S.~Mioduszewski$^{7,+}$}
\author{K.~Read$^{7}$}
\author{M.~Rosati$^{4}$}
\author{R.A.~Soltz$^{5}$}
\author{S.~Sorensen$^{7}$}
\author{J.H.~Thomas$^{5,\%}$}
\author{Y.~Torun$^{1,*}$}
\author{D.~Winter$^{2}$}
\author{W.A.~Zajc$^{2}$}

\address{$^{1}$Brookhaven National Laboratory, Upton, New York 11973}
\address{$^{2}$Columbia University, New York, NY 10027}
\address{$^{3}$Florida State University, Tallahassee, FL 32306}
\address{$^{4}$Iowa State University, Ames, IA 50010}
\address{$^{5}$Lawrence Livermore National Laboratory, Livermore, CA 94550}
\address{$^{6}$Oak Ridge National Laboratory, Oak Ridge, TN 37831}
\address{$^{7}$University of Tennessee, Knoxville, TN 37996}
\address{$^{\#}$Present address: University of Illinois at Urbana-Champaign, Urbana, IL 61801}
\address{$^{+}$Present address: Brookhaven National Laboratory, Upton, New York 11973}
\address{$^{\%}$Present address: Lawrence Berkeley National Laboratory, Berkeley, CA 94720} 
\address{$^{*}$Present address: Illinois Institute of Technology, Chicago, IL 60616}

\begin{abstract}
%
%

Published measurements of semi-inclusive \Lam\ production in \pAu\
collisions at the AGS are used to estimate the yields of singly
strange hadrons in nucleus-nucleus (A-A) collisions. Results of a
described extrapolation technique are shown and compared to
measurements of \Kp\ production in \SiAl, \SiAu, and \AuAu\ collisions
at the AGS and net \Lam\ production in \SuSu, \SAg, \PbPb, and inclusive
\pA\ collisions at the SPS.  The extrapolations can account for more
than 75\% of the measured strange particle yields in all of the
studied systems except for very central Au+Au collisions at the AGS
where RQMD comparisons suggest large re-scattering contributions.  

%
\end{abstract} \end{frontmatter}

Strange particle production has long been considered an important
experimental probe of heavy ion collisions due to the possibility that
strange particle yields may be significantly enhanced by quark-gluon
plasma (QGP) formation~\cite{Koc83:strange,Koc86:Probing}. Heavy ion
experiments at the Brookhaven National Laboratory AGS and CERN SPS
accelerators have observed factors of 3-4 enhancements in the yield per
participant of strange mesons and singly strange 
baryons~\cite{Abb90:Kaon,Ahle:1999va,Ahle:1998gv,Eis92:Rapidity,Ahm96:Lambda,Bar90:Production,Bachler:1992js,Alber:1994tz}
and, at the SPS, larger enhancements in the production of multiply
strange baryons~\cite{Antinori:1999dy,Andersen:1999ym,Appelshauser:1998va}.
Cascade models can quantitatively
reproduce the enhancement of strange mesons in central heavy ion
collisions at the AGS but are unable to reproduce the enhancements
observed at the SPS without the introduction of "exotic" processes.
Furthermore, inclusive measurements of kaon production in
proton-nucleus (\pA) collisions at the AGS have shown an enhancement
with target size~\cite{Abb91:Comparison,Abb92:Measurement}, while no
such enhancement was seen at the SPS~\cite{Bialkowska:1992kq}. The
strong enhancements in nucleus-nucleus (A-A) collisions at the SPS, 
combined with the
lack of enhancement in \pA\ and the failure of cascade models, have been
cited as experimental evidence for QGP formation at the
SPS~\cite{Heinz:2000ba,Heinz:2000bk}, while the enhancements at the AGS
are typically attributed to rescattering of produced hadrons. 
Then, the similarity in magnitude and pattern of enhancements of singly
strange hadrons at the AGS and SPS must be attributed to coincidence.

However, previous measurements of semi-inclusive \Lam\ production by
Experiment 910 at the AGS~\cite{Chemakin:2000ha} suggest an
alternative, common explanation for the enhancement of singly strange hadron
production in A-A collisions at the AGS and SPS.  E910 has
measured the dependence of the total \Lam\ yield in 17.5~GeV/c \pAu\
collisions on the number of scatterings \Nnu\ of the proton in the Au
nucleus and demonstrated enhanced \Lam\ production over a number of
participant (\Npart) scaling of \pp\ data. In this paper we present
the results of an analysis that extrapolates the effect observed by
E910 to A-A collisions and predicts the hyperon-associated \Kp\ yields
in \SiAl, \SiAu, and \AuAu\ collisions at the AGS and the net \Lam\
yields in \SuSu, \SAg, and \PbPb\ collisions at the SPS.  We show that
the extrapolation of the effect observed in p-A collisions can account
for most of the observed enhancements in hyperon-associated \Kp\ production at
the AGS and \Lam\ production at SPS in both light and heavy ion
induced reactions. With the same analysis we also explain the lack of
observed enhancement in inclusive \pA\ measurements at the SPS.

E910 reported~\cite{Chemakin:2000ha} that the \Lam\ excess in \pAu\ 
collisions over a simple \Npart\ or wounded-nucleon 
scaling of \pp\ data increased
proportionally to \Nnu\ for the first three collisions with a slope equal
to one half of the \pp\ \Lam\ multiplicity,
\begin{equation}
\Delta \NLam (\nu) = (\nu - 1) \half \NLamPP.
\label{eq:excess}
\end{equation}
Since the projectile contributes $\half \NLamPP$ to the total \Lam\
multiplicity in the first collision, the E910 excess is consistent with
a projectile contribution to the total \Lam\ multiplicity that grows
with \Nnu, as~\cite{Chemakin:2000ha}
\begin{equation}
N_\Lambda^{\rm proj} = \frac{1}{2} \NLamPP \, \nu, \nu \leq 3,
\label{eq:paprojlamnu}
\end{equation}
and remains constant for $\nu > 3$. To extrapolate to A-A
collisions we assume that Eq.~\ref{eq:paprojlamnu} is valid for all
relevant beam energies and that it applies equally to all
participating nucleons with \NLamPP\ replaced
by the estimated yield in isospin averaged nucleon-nucleon (\NN) collisions,
\Nnn. Then, the total multiplicity in an \AB\ collision is
\begin{equation}
\NAB = \frac{1}{2}\Nnn \sum_{\nu = 1}^{3} \left[ A P_{\rm p}(\nu) + B P_{\rm
t}(\nu) \right]  \nu  +  \frac{3}{2}\Nnn \sum_{\nu >3}
 A P_{\rm p}(\nu) + B P_{\rm t}(\nu).
\label{eq:yieldcalc}
\end{equation}
\PptNu\ represents the probability distribution for the projectile and
target nucleons to undergo \Nnu\ primary scatterings during the
collision. Since the number of participants can also be calculated
from the \PptNu, 
\begin{equation}
\Npart^{AB} = \sum_{\nu \geq 1} A P_{\rm p}(\nu) + B P_{\rm
t}(\nu),
\label{eq:npartcalc}
\end{equation}
we can directly obtain the relationship between \NAB\ and \Npart.  We
note that Eq.~\ref{eq:yieldcalc} applied to p-A collisions reduces
to
\begin{equation}
N^{\rm pA} = \Nnn \nu, \nu \leq 3 \; ; \;
N^{\rm pA} = \frac{1}{2} \Nnn (\nu + 3), \nu > 3. 
\label{eq:patotal}
\end{equation}
The total \Lam\ multiplicities measured by E910 saturate for $\Nnu >
5$ implying a violation of Eq.~\ref{eq:patotal}. This saturation
was attributed to the stopping of the projectile~\cite{Chemakin:2000ha} 
and is not expected to be present at SPS energies. 

In our analysis the \PptNu\ distributions were calculated using both a
Glauber model~\cite{Glauber:1970jm} and the 
Lund Monte-Carlo~\cite{And86:Fritiof}. 
The results are presented and compared in
Fig.~\ref{fig:bdependence} in terms of $\Rpart \equiv 2 \NAB/ (\Npart
\Nnn)$, {\it i.e.} the factor by which the A-A yield per participant is
enhanced relative to \NN\ collisions. In the Glauber calculation, the
\PptNu\ distributions were evaluated at fixed values of collision
impact parameter $b$, and the simultaneous dependence of \NAB\ and
\Npart\ on $b$ was used to obtain $\NAB(\Npart)$.  This procedure,
while calculationally simple, potentially introduces errors by
averaging over fluctuations that might differently affect \NAB\ and
\Npart.  The slight differences between the two calculations at large
$b$ are consistent with the expectation that the Monte-Carlo
calculation would better account for these fluctuations. 
The two calculations give very similar results for the \Npart\ dependence of
\Rpart, though the Monte-Carlo result has a slightly slower growth in
\Rpart\ with \Npart. 
To evaluate the importance of the stopping effects at AGS energies
implied by the E910 data, we additionally 
imposed in the Glauber calculations a maximum
thickness to the nuclei corresponding to 5 nucleon interaction
lengths with the result shown in Fig.~\ref{fig:bdependence} by the
dashed lines.  
Hereafter we will use the Monte-Carlo results when
presenting the calculation results. 

We now compare our calculations to experimental data starting with \Kp\
multiplicities measured by Experiments 859 and 866 in \SiAl, \SiAu, and
\AuAu\ collisions at the AGS~\cite{Ahle:1999va}. At AGS energies, the \Kp\
is predominantly produced in association with hyperons, so 
the total \Kp\ multiplicity should have the same \Nnu\ dependence as the
\Lam\ (excluding \KpKbar\ contributions).  Previously estimated
production rates for \Kp\ production in isospin-averaged \NN\
collisions are listed in Table~\ref{tbl:pprates} for the three
different colliding systems~\cite{Ahle:1999va}.  At comparable energies 
\pp\ measurements~\cite{Fes79:Strangeness} suggest that 15\% of the
\Kp\ yield results from \KpKbar\ processes.  Lacking clear knowledge of how
these processes grow in \pA\ collisions we conservatively assumed that
\KpKbar\ production increases proportionally to \Npart.  Thus, the 
\NN\ ~\Kp\ rates used in Eq.~\ref{eq:paprojlamnu} were reduced by 15\%, and
the \KpKbar\ contribution to the \Kp\ yield was added separately.
Figure~\ref{fig:agsdata} shows a comparison between the results of
our calculation and the experimentally measured \Kp\ yields.  Two sets
of points are shown for the \SiAu\ data because the experimentally
quoted \Npart\ values were found to be inconsistent with calculated
values of \Npart\ for the quoted centrality fractions~\cite{Ahle:1999va}, 
though nowhere by more than twice the stated error on the \Npart\ values.  
Since the
two calculations give results that differ by less than 2\%, we
attribute the discrepancy with the data to a systematic error in the
method used by the experiment to estimate \Npart\ in asymmetric
collisions.  The shaded region in Fig.~\ref{fig:agsdata} indicates the
uncertainties in our extrapolation due to the systematic errors on the \NN\
yields shown in Table~\ref{tbl:pprates} and the uncertainty in the
\KpKbar\ rate in \pp\ collisions.  Our calculations account for more
than 75\% of the measured \Kp\ yields everywhere except in the most
central \AuAu\ collisions where the data exceeds our extrapolation by
50\%.  We also show in
Fig.~\ref{fig:agsdata} the results from the cascade model 
RQMD~\cite{Sor95} simulating 11.1~AGeV/c Au+Au collisions.  Although 
RQMD significantly under-predicts the E866 \Kp\ yield 
over most of the centrality range, it shows a rapid increase in
\Kp\ yield with \Npart\ for very central collisions that may be due to
rescattering processes, which would be expected to grow roughly
like $\Npart^2$.  The rapid rise in the RQMD calculation is qualitatively
similar in shape to the observed excess of the central \AuAu\ data
over our calculation suggesting that this excess may, in fact, be due
to rescattering processes not present in p-A data. 

We also apply our extrapolation procedure to SPS energies to
determine whether the strangeness enhancements
there may also be accounted
for by the effect observed in the E910 data. Since
Eq.~\ref{eq:paprojlamnu} may not describe baryon-anti-baron (\BBbar)
production which is negligible at AGS energies but important at SPS
energies, comparisons with SPS data are done using the ``net''
($N_\Lambda - N_{\bar{\Lambda}}$) yields.  Figure~\ref{fig:spsdata} shows
measurements of total net \Lam\ production from NA35 in central 
\SuSu\ \cite{Bar90:Production} and \SAg\ \cite{Alber:1994tz} collisions at
200~\AGeV\ and net \Lam\ production at mid-rapidity in \PbPb\
collisions at 160~\AGeV\ from WA97~\cite{Andersen:1999ym}. The input
\NN\ rates required for the extrapolation are listed in
Table~\ref{tbl:pprates}. For the \PbPb\ extrapolation we reduced the
\pBe\ \Lam\ yields of WA97 by a factor of 0.87 to remove the
enhancement that would result from multiple collisions of the 
proton according to our method.  The extrapolations, shown in 
Fig.~\ref{fig:spsdata} by solid lines, reproduce the 
observed \Lam\ enhancement in \SA\ collisions but
fall well below the \PbPb\ data.  
However, the problem with the latter comparison is that the WA97 measurements
are limited to an acceptance of unit rapidity about the \NN\ center of mass.
The \pp\ \Lam\ rapidity distribution is
strongly peaked at forward/backward rapidity and suppressed near
mid-rapidity~\cite{Brick:1980vj,Allday:1988bb}. The increased stopping
of the baryons in \PbPb\ collisions~\cite{Appelshauser:1998yb}
will strongly shift the peaks toward mid-rapidity, so that an
enhancement would be observed by WA97 even if there were no
enhancement in the {\em total} \Lam\ yield. From  parameterizations
of the \Lam\ $x_F$ distribution in \pBe\ collisions~\cite{Kachelhoffer:1997qs}
we estimate that 11\% of the total
\pBe\ \Lam\ yield is contained within the central unit of rapidity.
Using recent measurements and extrapolations for \Lam\ and \Lambar\
\dndy\ for 158~\AGeV\ Au-Au~\cite{Anticic:2004xx} we estimate that
21\% of the total net \Lam\ yield is contained in the central unit of
rapidity. If we multiply our 
extrapolation by a factor of $0.21/0.11=1.9$ to account for the
effects of the \dndy\ shape change, we then obtain the dashed line shown in 
Fig.~\ref{fig:spsdata}. The excellent agreement between this
result and the \PbPb\ data may be partially accidental since we have
not included an \Npart\ dependence to the correction factor. 
Nonetheless, we can conclude that our extrapolation accounts for more 
than 75\% of the net hyperon yields 
in light and heavy ion collisions at the SPS.

The large enhancements at SPS energies predicted by the E910 extrapolation 
appear to contradict the lack of enhancement 
noted by previous \pA\ measurements at the 
SPS~\cite{Bialkowska:1992kq}.  However, the effect observed by
E910 may be substantially obscured in inclusive collisions due to the
large contributions from peripheral collisions and due to the
saturation of the projectile enhancement at $\nu =3$, which will
result in a decreasing \Lam\ yield per participant for $\nu>3$. 
Figure~\ref{fig:pasps} shows results of our calculations for 
inclusive collisions
compared with a variety of inclusive measurements~\cite{Bialkowska:1992kq}.  
Our calculation reproduces the nearly 
constant \Lam\ yield per participant observed in the experimental data
in spite of the enhanced projectile contribution. 

In conclusion, we have performed a set of calculations intended to
quantitatively evaluate the significance of the observed enhancement of
\Lam\ production in \pAu\ collisions~\cite{Chemakin:2000ha} in
the interpretation of 
strangeness enhancement in heavy ion collisions. Our extrapolation is
based on applying the enhancement observed by E910 to all participant
nucleons in A-A collisions assuming that it results directly from the
multiple interactions of the proton. We show that our extrapolation can
account for more than 75\% of the total hyperon or hyperon associated
kaon yield in light and heavy ion induced collisions at the AGS and
SPS {\em except} for \Kp\ production in central \AuAu\ collisions.
Here, comparisons with RQMD suggests that the discrepancy is due to
rescattering processes that contribute most importantly in central
collisions. 
The analysis presented here
suggests that a dynamical mechanism and not QGP formation may be
responsible for enhancing the yields of (at least) singly strange
hadrons. 
Our analysis also suggests that the
similar enhancements observed at the AGS and SPS result from a common
mechanism in spite of the difference in beam energies.

\bibliography{aacomp}
\bibliographystyle{prsty}

\begin{table}
\begin{tabular}{|r|c|c|r|r|l|}\hline
Beam p (GeV/c) & System & Part. & Mult. & Error & Ref. \\ \hline \hline
14.6 & Si+A  & \Kp\     & 0.049 & 0.008 & \cite{Ahle:1999va} \\
14.6 & Si+A  & \KpKbar\ & 0.007 & 0.002 & \cite{Fes79:Strangeness} \\
11.1 & Au+Au & \Kp\     & 0.033 & 0.005 & \cite{Ahle:1999va} \\
11.1 & Au+Au & \KpKbar\ & 0.005 & 0.002 & \cite{Fes79:Strangeness} \\
200  & S+A   & \Lam\    & 0.10  & 0.01  & \cite{Gazdzicki:1996pk} \\
200  & S+A   & \Lambar\ & 0.013 & 0.005 & \cite{Gazdzicki:1996pk} \\
160  & Pb+Pb & \Lam\    & 0.0334 & 0.0005 & \cite{Antinori:1999jd} \\
160  & Pb+Pb & \Lambar\ & 0.0111 & 0.0002 & \cite{Antinori:1999jd} \\ \hline
\end{tabular}
\caption{Measured or estimated rates for \Kp, \KpKbar, \Lam, and \Lambar\ production
in nucleon-nucleon collisions used to extrapolate to corresponding
measurements in indicated light/heavy ion collisions. }
\label{tbl:pprates}
\end{table}

\begin{figure}[h]
\caption{Calculated enhancement factor \Rpart\ (see text for details) vs $b$ and \Npart\ for various colliding systems, solid lines - Glauber geometry, points - Lund geometry.}
\label{fig:bdependence}
\end{figure}

\begin{figure}[h]
\caption{Comparison between measured \Kp\ yields in \SiA\ (left) and \AuAu\
collisions (right) at the AGS and extrapolations, $\blacksquare$ -
\SiAl, $\triangle$ - \SiAu, $\bullet$ - \AuAu, $\blacktriangledown$ -
\SiAu\ with adjusted \Npart, solid line - extrapolations, dashed -
\Npart\ scaling of \NN, dot-dashed - RQMD calculation for
\AuAu. Shaded regions indicate $1 \sigma$ systematic errors due to
uncertainties in \NN\ yields.} 
\label{fig:agsdata}
\end{figure}

\begin{figure}[h]
\caption{
Comparison of net \Lam\ (\LamminusLambar) yields from (left) \SA\ and (right) \PbPb\ collisions at the SPS and calculations, $\blacktriangledown$ - \SuSu, $\blacktriangle$ - \SAg, $\bullet$ - \PbPb. Lines, solid - extrapolations, dashed - \Npart\ scaling of \NN, dot-dashed (right) extrapolation corrected for stopping (see text for details)
}
\label{fig:spsdata}
\end{figure}

\begin{figure}[h]
\caption{
Comparison of average net \Lam\ yield per participant in inclusive
\pA\ collisions 200 GeV/c and extrapolation of E910 data, $\bullet$ - 
\pA, $\square$ - \pp, solid line - extrapolation. 
}
\label{fig:pasps}
\end{figure}

\begin{figure}[h]
\centerline
{
\psfig{file=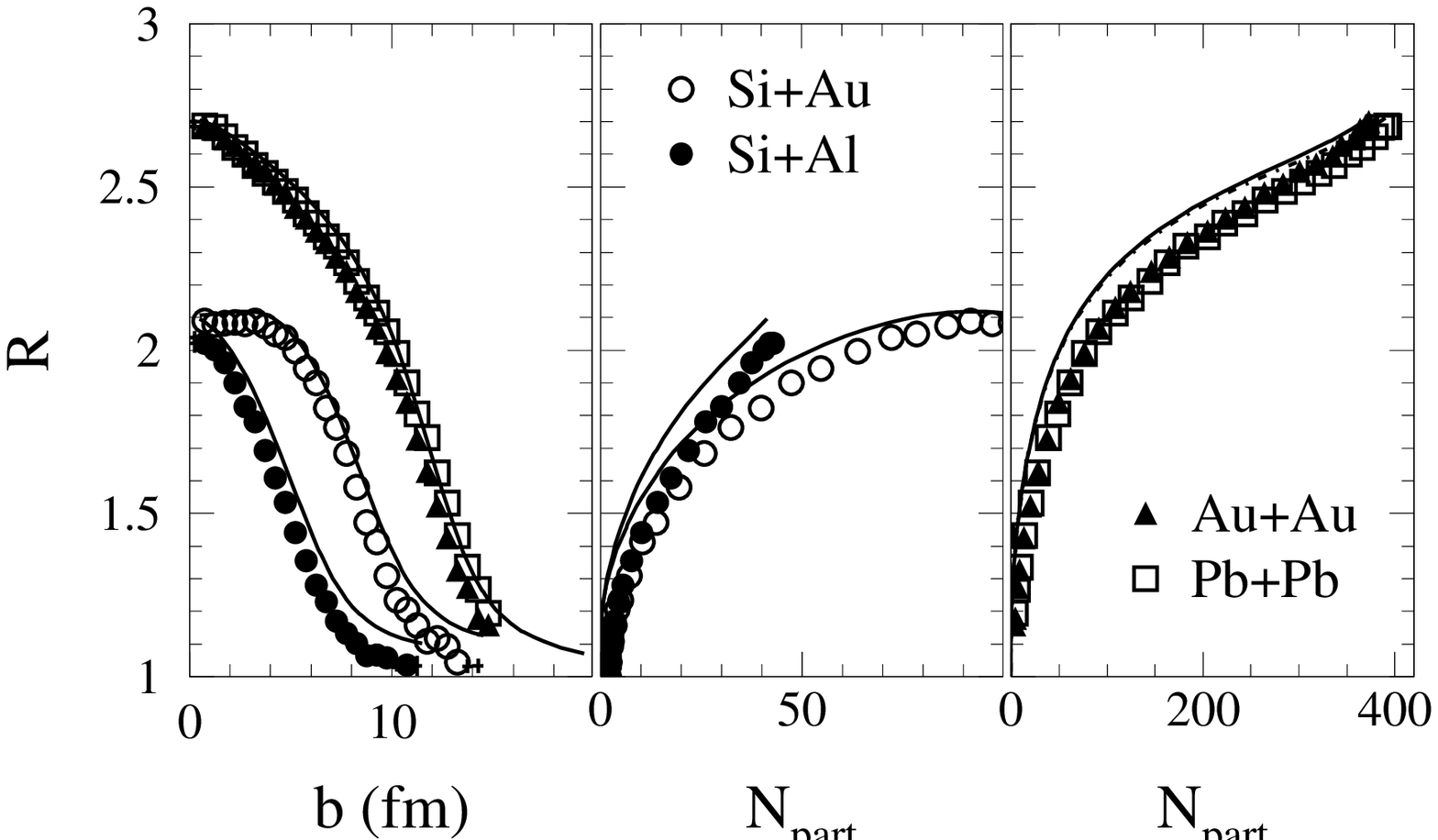,height=3.5in,bbllx=0pt,bburx=560pt,bblly=0pt,bbury=320pt}
}
\end{figure}

\begin{figure}[h]
\centerline
{
\psfig{file=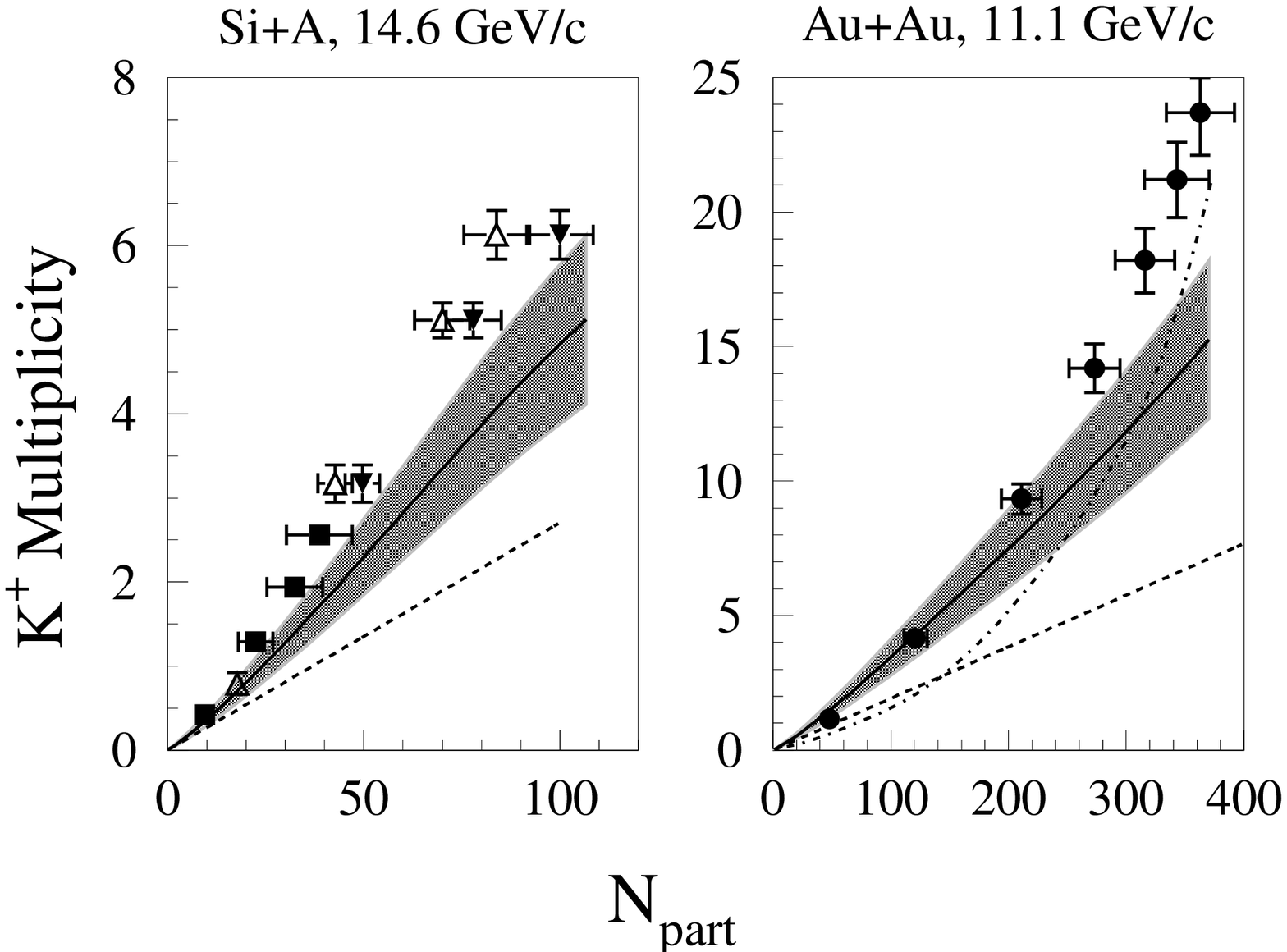,height=3.5in,bbllx=0pt,bburx=560pt,bblly=0pt,bbury=420pt}
}
\end{figure}

\begin{figure}[h]
\centerline
{
\psfig{file=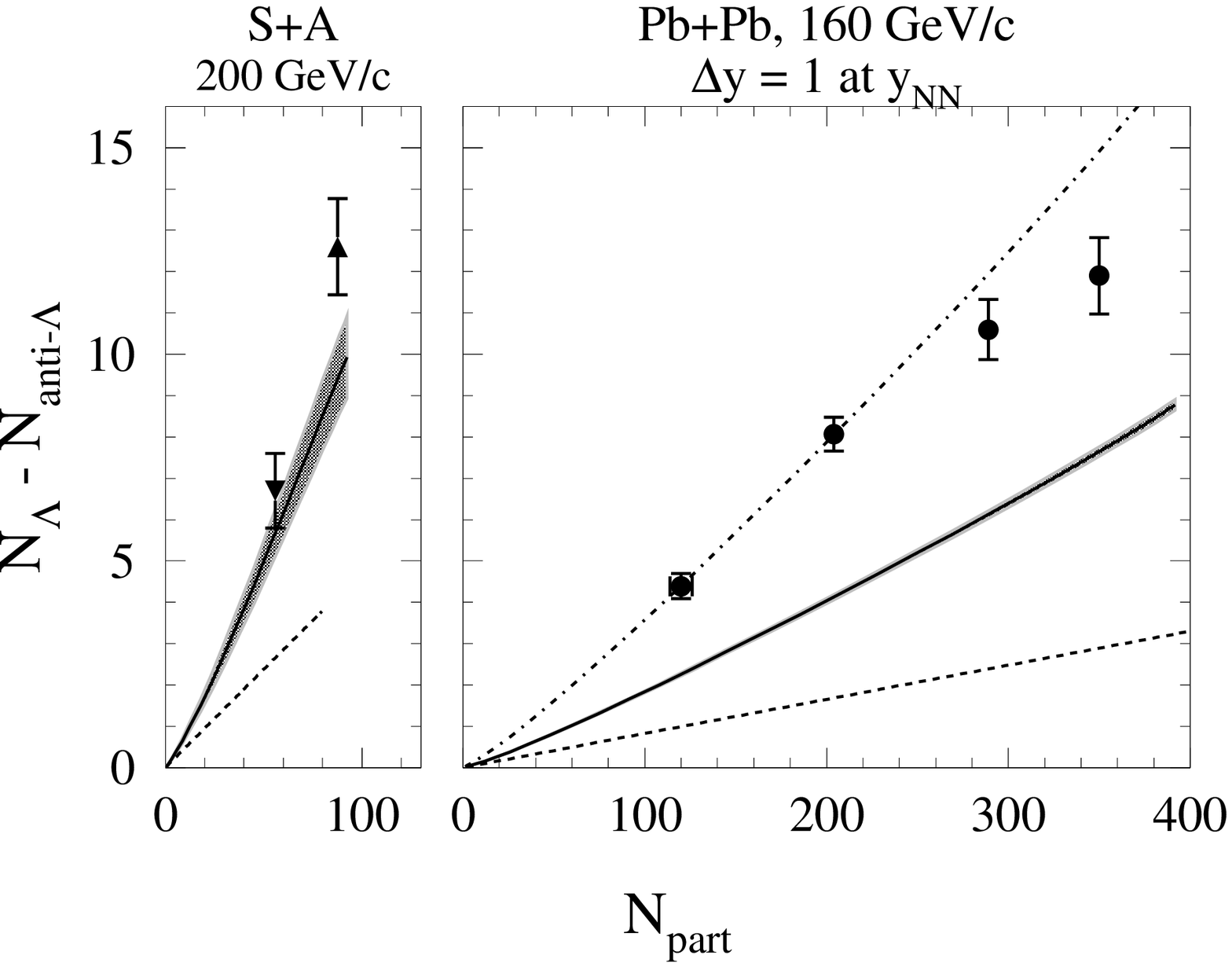,height=3.5in,bbllx=0pt,bburx=560pt,bblly=0pt,bbury=430pt}
}
\end{figure}

\begin{figure}[h]
\centerline
{
\psfig{file=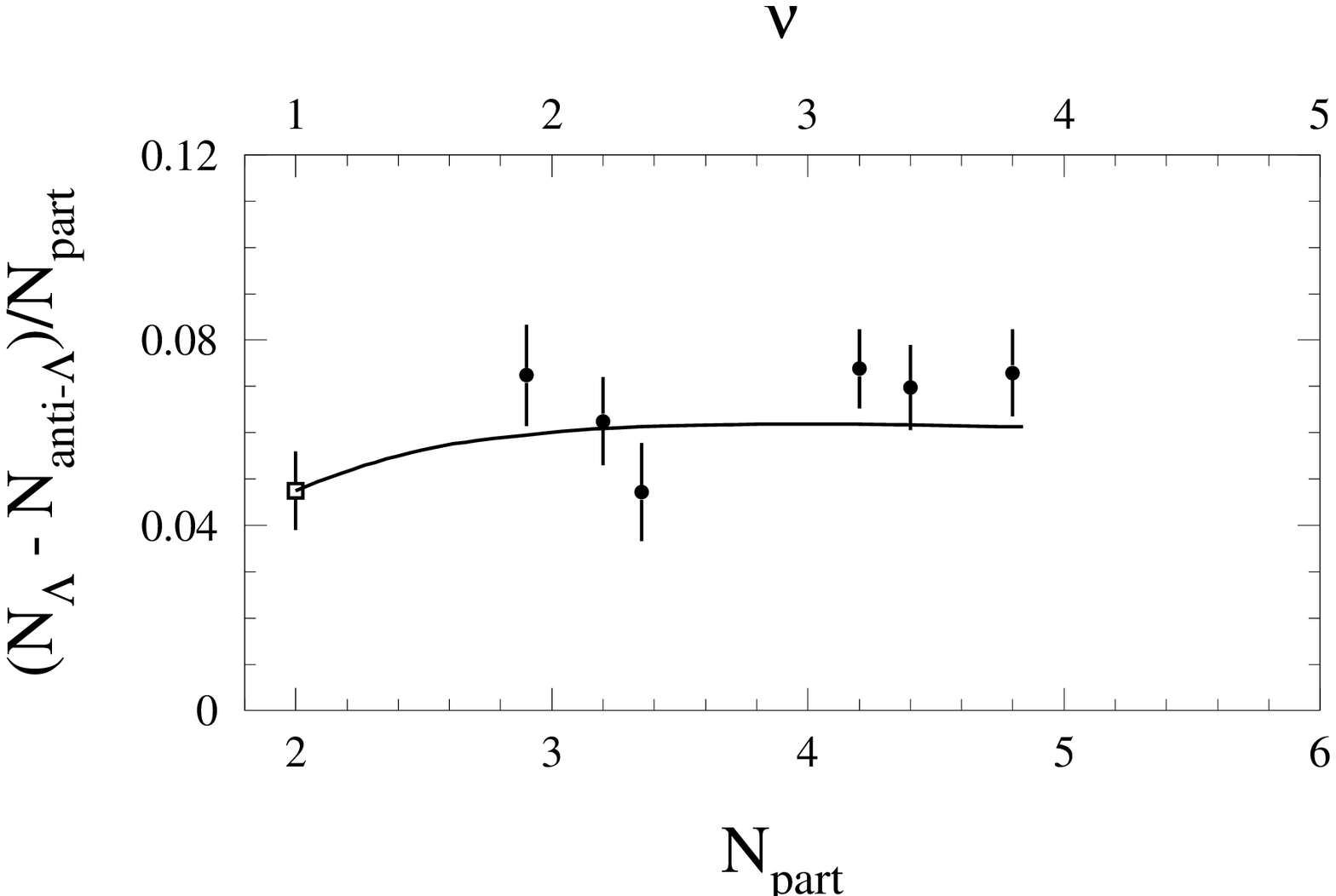,height=2.5in,bbllx=0pt,bburx=560pt,bblly=0pt,bbury=355pt}
}
\end{figure}

\end{document}